\documentclass[aps,pre,twocolumn,superscriptaddress,floatfix]{revtex4}
\usepackage{dcolumn}
\usepackage{amsmath}
\usepackage{amssymb}
\usepackage{graphicx}
\usepackage{times}
\usepackage{bm}
\usepackage[T1]{fontenc}
\usepackage{color}
\usepackage{url}
\usepackage[bf]{subfigure}
\usepackage{rotating}

\begin{document}

\title{A machine learning approach to predicting dynamical observables from network structure}

\author{Francisco A. Rodrigues}
\affiliation{Mathematics Institute, University of Warwick, Gibbet Hill Road, Coventry CV4 7AL, UK.}
\affiliation{Centre for Complexity Science, University of Warwick, Coventry CV4 7AL, UK.}
\affiliation{Instituto de Ci\^{e}ncias Matem\'{a}ticas e de Computa\c{c}\~{a}o, Universidade de S\~{a}o Paulo, S\~{a}o Carlos, SP, Brazil.}

\author{Thomas Peron}
\affiliation{Instituto de Ci\^{e}ncias Matem\'{a}ticas e de Computa\c{c}\~{a}o, Universidade de S\~{a}o Paulo - Campus de S\~{a}o Carlos, Caixa Postal 668, 13560-970 S\~{a}o Carlos, SP, Brazil.}

\author{Colm Connaughton}
\affiliation{Mathematics Institute, University of Warwick, Gibbet Hill Road, Coventry CV4 7AL, UK.}
\affiliation{Centre for Complexity Science, University of Warwick, Coventry CV4 7AL, UK.}

\author{J\"{u}rgen Kurths}
\affiliation{Potsdam Institute for Climate Impact Research, D14473
Potsdam, Germany}
\affiliation{Department of Physics, Humboldt University, 12489 Berlin, Germany}

\author{Yamir Moreno}
\affiliation{Institute for Biocomputation and Physics of Complex Systems (BIFI), University of Zaragoza, 50018 Zaragoza, Spain}
\affiliation{Department of Theoretical Physics, University of Zaragoza, 50018 Zaragoza, Spain}
\affiliation{ISI Foundation, Turin, Italy}

\begin{abstract}
Estimating the outcome of a given dynamical process from structural features is a key unsolved challenge in network science. The goal is hindered by difficulties associated to nonlinearities, correlations and feedbacks between the structure and dynamics of complex systems. In this work, we develop an approach based on machine learning algorithms that is shown to provide an answer to the previous challenge. Specifically, we show that it is possible to estimate the outbreak size of a disease starting from a single node as well as the degree of synchronicity of a system made up of Kuramoto oscillators. In doing so, we show which topological features of the network are key for this estimation, and provide a rank of the importance of network metrics with higher accuracy than previously done. Our approach is general and can be applied to any dynamical process running on top of complex networks. Likewise, our work constitutes an important step towards the application of machine learning methods to unravel dynamical patterns emerging in complex networked systems.

\end{abstract}

\maketitle

Modern network science has been successful at showing that accounting properly for the interaction patterns of a system's components is crucial for describing its functionality~\cite{Boccaletti06,Barrat08,Pastor015,Rodrigues016,Arruda018}. Admittedly, tackling the laws that govern the relationship between the structure and function of a system is a formidable challenge. Achieving such a goal would entail not only being able to assess the impact of some structural patterns in the dynamics of networks, but also to predict dynamical outcomes from an incomplete, and often noisy, knowledge of the structure of a system. The problem is not a minor one: going from knowing the structure of a system to anticipating its dynamical response implies to sort out nonlinearities in the nodes' responses, spatial and temporal correlations due to complex interconnection patterns, and feedbacks resulting from these interactions, among other difficulties. 

On the other hand, it is known that not all network properties impact equally the dynamics of the system. Some properties are more important than others. For instance, in the context of disease spreading, the degree distribution is crucial for the critical properties of the system $-$that is, whether there is a vanishing threshold or not$-$, whereas correlations play a less determinant role~\cite{Satorras015, Arruda018}. Similar conclusions can be extracted for other dynamical processes, such as rumor spreading~\cite{Kitsak010, Arruda013} and synchronization phenomena ~\cite{Arenas08,Rodrigues016}. Tracking down which network properties are fundamental for the dynamics of the system will not only allow for more accurate dynamical predictions, but also to provide actionable responses to topological changes or control interventions to drive the system to a desired global outcome. Applications range from identifying influential and core spreaders in disease dynamics to tuning the level of synchronization of power grids, electronic circuits and neuronal systems. 

Here, we address the two challenges described above. On the one hand, we develop a methodology that allows predicting several macroscopic observables for two paradigmatic dynamics: disease contagion and synchronization. In doing so, we also report on what are the most determinant topological properties for such a prediction to be accurate. Specifically, we present a general approach to predict variables associated to dynamical processes in complex networks, namely, the level of synchronization of Kuramoto oscillators and the outbreak size in a SIR (susceptible-infected-recovered) dynamics. We also evaluate the importance of network properties for the prediction of such dynamical variables. Our method is general and could be applied to other scenarios in which the aim is to predict a random dynamic variable from a subset of nodes and their dynamics. The methodology proposed here opens new paths to investigate the structure and dynamics of complex systems through modern methods of machine learning.

To formulate our framework, we depart from traditional methods from nonlinear dynamics and statistical mechanics~\cite{Mehta018} and instead use a machine learning (ML) approach. To specify the predictive learning model, we associate a dynamic variable $Y_i$ to each node $i$. This variable is time dependent, but can reach a constant value for stationary processes, $Y_i(t \rightarrow \infty) = Y_i$. In the case of synchronization, $Y_i$ can be an indicator variable such that $Y_i=1$ if oscillator $i$ is synchronized and 0 otherwise. In the case of disease spreading, we define $Y_i$ as the expected fraction of infected nodes when the disease is seeded at node $i$, i.e.,
\begin{equation} \label{Eq:Yi-epidemics}
Y_i = \lim_{t \rightarrow \infty} \frac{1}{N} \sum_{j=1}^N Z_j(t),
\end{equation}
where $Z_j=1$ if node $j$ is infected and $Z_j=0$, otherwise. The variable $Y_i$ depends on the network structure, because the state of each node $i$ is defined by the interaction with its neighbors. Thus, one can assume that $Y_i$ is related to a feature vector extracted from the network structure, $\mathbf{X_i} = \{X_{i1}, X_{i2}, \ldots, X_{id}\}$, as
\begin{equation} \label{Eq:y}
Y_i = f(\mathbf{X_i}) + \epsilon,
\end{equation}
where $N$ is the number of nodes and $\epsilon$ is a random error term independent of $\mathbf{X_i}$ and normally distributed with mean zero and standard deviation $\sigma$. The feature vector represents measures that describe structural properties, which should be carefully chosen to predict the variable $Y_i$ with as higher accuracy as possible. Although the selection of the elements in $\mathbf{X_i}$ can be done using methods for model comparison and feature selection algorithms, here we consider traditional network metrics. Specifically, we include (i) degree ($K$), (ii) clustering coefficient ($C$), (iii) closeness centrality ($CC$), (iv) betweenness centrality ($B$), (v) eigenvector centrality ($EC$), (vi) Page Rank ($PR$) and (vii) k-core ($KC$). Note that, except for the clustering coefficient, we consider measures that essentially quantify the centrality of the nodes from different perpectives~\cite{Rodrigues018}. 

The fixed but unknown function $f : \Re^d \rightarrow \Re$ in Eq~\eqref{Eq:y} represents a surface in a $d$-dimensional space, where $d$ is the number of network measures in $\mathbf{X_i}$. Its estimation is a complex high-dimensional problem that cannot be solved using traditional statistical methods because the observations are not independent. On the other hand, in terms of statistical learning, $f$ can be estimated by inference or prediction~\cite{James013, Mehta018}. In the case of inference, the goal is to understand the relationship between $Y_i$ and $\mathbf{X_i}$, determining, for instance, the contribution of each measure $j$, $X_{ij}$, to the dependent variable $Y_i$. This kind of analysis has been applied to study the synchronization of Kuramoto oscillators using linear regression~\cite{Arruda013}. On the contrary, prediction aims at finding the estimation $\widehat{Y_i}$ that can best predict $Y_i$ accurately~\cite{James013}. Here we are interested in this latter approach, namely, in predicting the expected number of infected nodes when the disease starts at any given node $i$ or the state of oscillators for the synchronization dynamics. Note that the estimation of $Y_i$ is a regression problem for the disease case since $\{Y_i \in \Re | 0 \leq Y_i \leq 1\}$, $i=1, \ldots, N$, whereas it is a classification problem for the synchronization dynamics, given that in such a case the goal is to classify oscillators as partially phase-locked or unlocked.

As we are not interested in the exact form of the function $f$, but in optimum prediction of $Y_i$, we consider a ML method for the estimation of the statistical model given by Eq~\eqref{Eq:y}. To this end, we need to specify a set of training examples~\cite{Mehta018}, so that the machine learning algorithm learns from data automatically~\cite{Theodoridis, Bishop06}, as well as a loss function, e.g., the minimization of the mean square error, $E = [Y_i - \widehat{f}(X_i)]^2/n$~\cite{Theodoridis}, where $\widehat{f}(X_i) \approx Y_i$. In what follows, we show results obtained for the two dynamical processes mentioned before applying two ML algorithms, namely, artificial neural networks and random forests~\cite{James013, Theodoridis08}. We stress again that our interest is in proposing a methodology that addresses the structure-dynamics dependency on networks, instead of a comparison of what is the most accurate approach one could adopt. 

\begin{figure}[!t] 
\begin{center}
\includegraphics[width=.99\columnwidth]{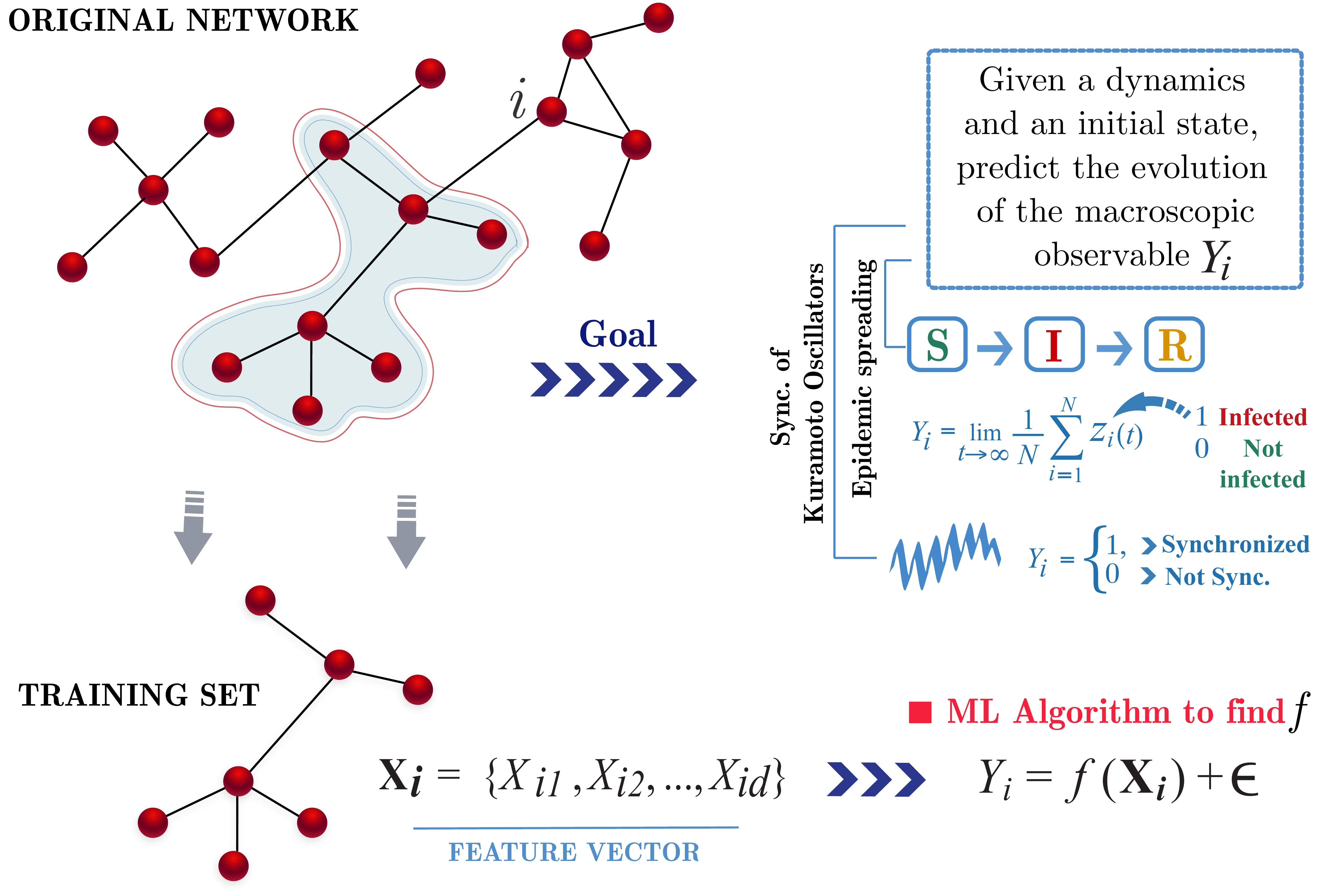}
\end{center}
\caption{Schematic representation of the methodology employed in this paper. A subset of nodes of a given network is used to train a ML algorithm that makes predictions about the dynamics of the entire system.}
\label{fig:scheme}
\end{figure}

To perform the classification and regression, we define a set of nodes in the training set, used to adjust the function $f$ in Eq.~\eqref{Eq:y}. The remaining vertices that do not belong to the training set are used to evaluate the accuracy of the prediction. The way in which the training and testing sets are selected depends on the dynamics. For the prediction of the outbreak size, we consider the $k$-fold cross validation. This implies that the original data set is randomly partitioned into $k$ equally-sized sets, where $k-1$ sets are used to train the algorithm and the remaining one is used for testing. The cross-validation process is then repeated $k$ times, with each of the $k$ sets used exactly once as the validation data. For the synchronization dynamics, being it a classification problem, we implement a stratified $k$-folds cross-validator with $k=10$ folds.  The stratified $k$-folds cross-validator returns stratified folds, preserving the percentage of samples for each class. The general approach developed here is summarized in Fig.~\ref{fig:scheme} and the classifier and regression models are available at the Scikit-learn library~\cite{scikit-learn}. Next, we present our results for the two dynamical processes here considered.

\paragraph*{Epidemic spreading} The prediction of when, how and to which extent an epidemic outbreak will take place is one of the most important challenges of modern science with fundamental implications for public health~\cite{Keeling08}. For instance, the prediction of the expected number of cases once a disease is seeded in a given individual or group of subjects would enable the identification of the most influential spreaders~\cite{Kitsak010, Arruda014} and contribute to develop methods for an efficient vaccination and disease control. Specifically, we consider a SIR (susceptible-infected-recovered) model whose dynamics has an absorbing state, i.e., the number of infected nodes goes to zero in a finite time~\cite{Keeling08, Pastor015}. We conjecture that the outbreak size, $Y_i$ (Eq.~\eqref{Eq:Yi-epidemics}), depends on the topological properties as given by Eq.~\eqref{Eq:y}, where we consider that the feature vector $\mathbf{X_i}$ contains the structural metrics listed before (i.e., properties (i) through (vii)). To determine the form of $f$ in Eq.~\eqref{Eq:y}, we use both artificial neural networks and random forests and predict the average size of outbreaks starting from each node $i$. 

\begin{figure}[!t] 
\begin{center}
\includegraphics[width=.9\linewidth]{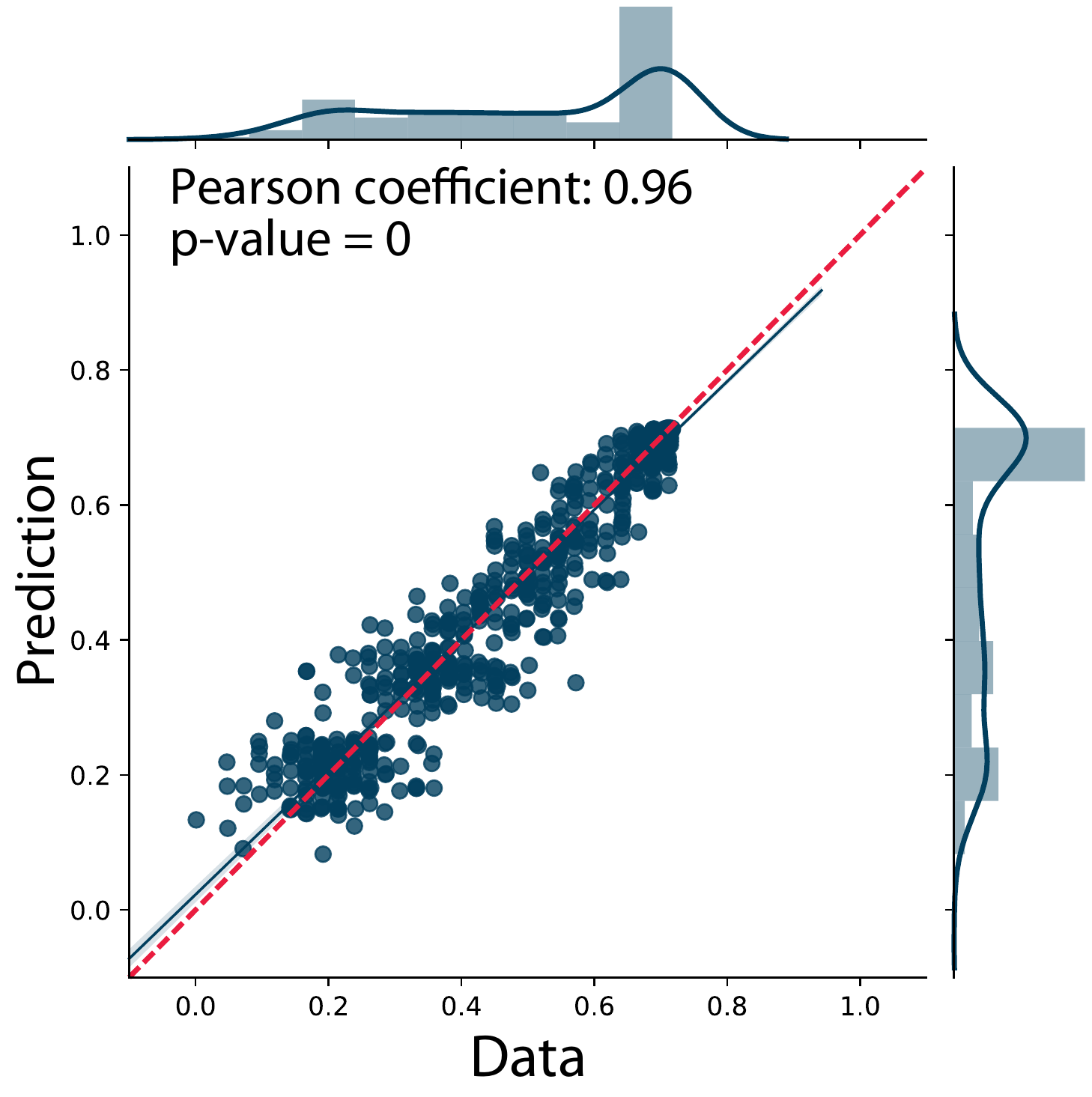}
\end{center}
\caption{Prediction of the outbreak size in terms of network structure for the US air transportation network. The SIR model is simulated using $\beta = 0.3$ and $\mu = 1$. The fitted regression line is represented by the straight line, whereas the dashed line represents $x=y$. The associated coefficient of determination for the neural network is $R^2 = 0.94$ for the airport network.}
\label{fig:prediction}
\end{figure}

To validate our approach, we use spreading data obtained from Monte Carlo simulations of the model with discrete time~\cite{Gomez010}. The process starts at one seeded node $i$, $i=1,\ldots, N$, and at each time step, each infected node contacts all its neighbors and transmits the disease with probability $\beta$. Infected nodes recover with probability $\mu$ and the dynamics stops when there are no infected nodes in the network. In our simulations, we consider $\mu = 1$ and $\beta  = 0.3$. The training step is performed with $50\%$ of the nodes, whereas the test is evaluated on the remaining set. Note that the observations used for testing are not considered in the training phase. Besides, we chose at random the training set, 
 but other methods are possible too (e.g., use of a random walk or preferential sampling according to the degree). Figure~\ref{fig:prediction} shows the outbreak size predicted by the model, Eq.~\eqref{Eq:y}, for the US air transportation network ($N=1,574$ airports and $\langle k\rangle = 35.87$ edges per airport). We have also tested the methodology for the Hamsterster social network ($N=1,788$ users and $\langle k\rangle = 13.95$  edges per user)~\cite{hamsterster}, obtaining a similar fitting (see also Fig.~\ref{fig:importance-epidemics} below). The result shown corresponds to the artificial neural network (we used a multilayer perceptron with 10 layers and 10 neurons in each layer), but the random forest provides similar outcomes. As it can be seen from the figure~\ref{fig:prediction}, our findings reveal that the information contained in the vector $\mathbf{X_i}$ is suitable to predict $Y_i$ with high accuracy.

Although both methods provide similar prediction errors, the random forest presents some advantages compared to neural networks. For instance, it has only a few parameters to be tuned, like the number of trees (we used 100 trees), contrariwise to the neural network, in which we have to set the number of neurons in each layer, the number of layers, the activation function and the solver for weight optimization~\cite{scikit-learn}. Important enough, a random forest algorithm also enables the quantification of the features' importance.  Basically, the importance is computed by measuring how effective the feature is at reducing uncertainty when creating the decision trees. All the values provided by the random forest sum to one and give the percentage of contribution of each measure on the prediction of the estimated quantity. Figure~\ref{fig:importance-epidemics} displays the results obtained when the estimation of the outbreak size is broken down as a function of the contribution of each of the network metrics used in the feature vector. For both the US airport and Hamsterster networks, the $k$-core is the most determinant feature to predict the outbreak size. This result agrees with previous analysis of influential spreaders in networks~\cite{Kitsak010, Arruda014}. However, note that our method is more general, since the results show that there is no unique measure to identify such main propagators, but a combination of them. Moreover, observe that the contribution (influence) of each metric strongly depends on the kind of network. For instance, for the airport network, the degree is almost as relevant as the $k$-core, which is not the case for the social network analyzed here. It is also worth remarking that, at variance with previous works~\cite{Kitsak010}, our methodology allows for a comparison of the influence of as many network properties as desired concurrently, i.e., when their possible role is considered altogether and they are combined to make the prediction better. Admittedly, as seen in the inset of Fig.~\ref{fig:importance-epidemics}, although the $k$-core is the most important feature, it alone is not enough to predict with the same accuracy the disease's outbreak size, Fig.~\ref{fig:prediction}.

\begin{figure}[!t]
\begin{center}
\includegraphics[width=0.95\columnwidth]{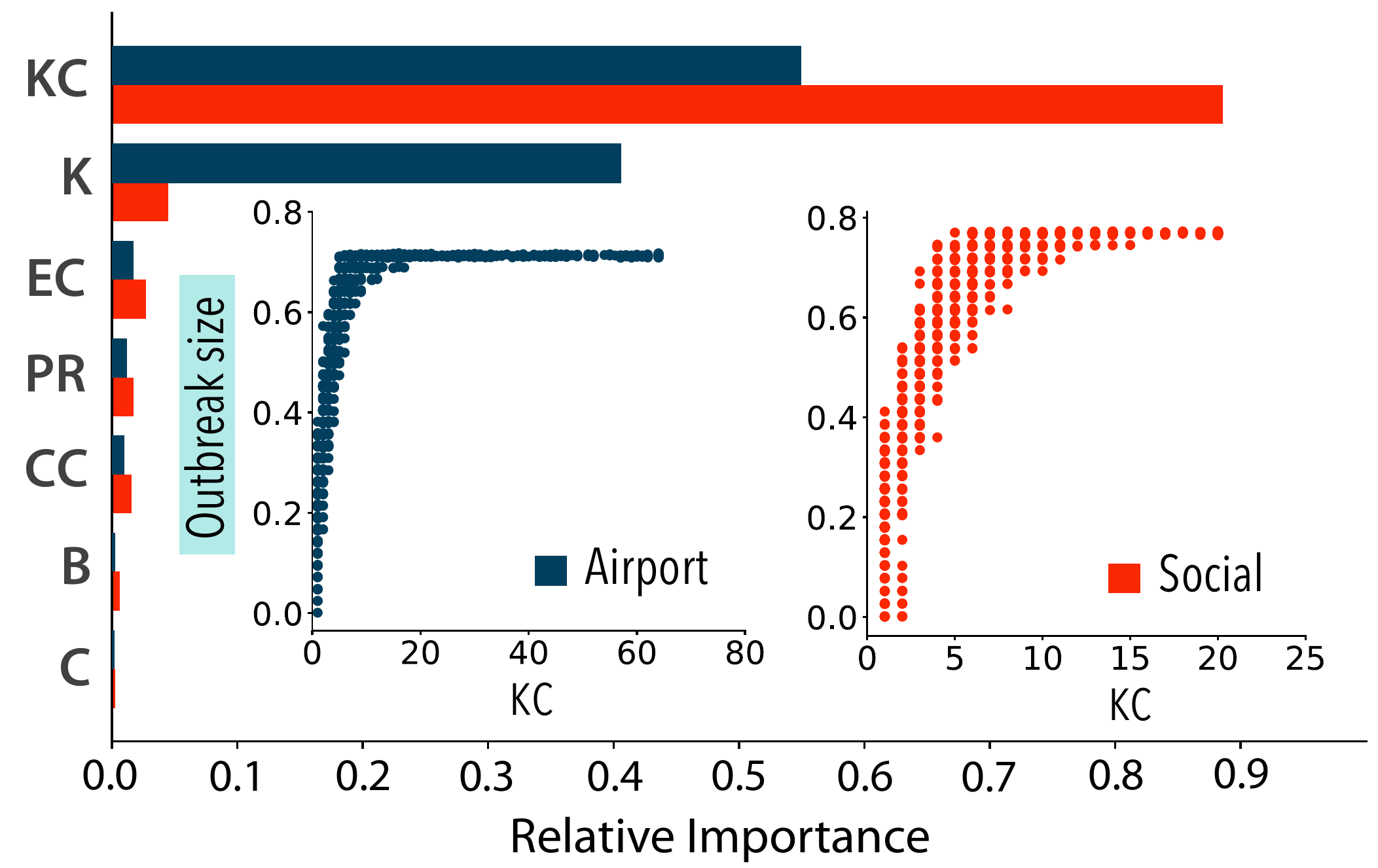}
\end{center}
\caption{Importance of network measures on the prediction of the outbreak size for the US air transportation and the Hamsterster social networks. The inset shows the most important measure and the respective outbreak size. The network measures are the degree ($K$), clustering coefficient ($C$), closeness centrality ($CC$), betweenness centrality  ($B$), eigenvector centrality ($EC$), Page Rank ($PR$) and k-core ($KC$). }
\label{fig:importance-epidemics}
\end{figure}

\paragraph*{Synchronization} The Kuramoto model (KM) is a paradigmatic model to study synchronization phenomena~\cite{Rodrigues016}. In complex networks, the KM is described by the set of equations,
\begin{equation}
\frac{d\theta_i}{dt} = \omega_i + \lambda \sum_{i=1}^N A_{ij}\sin(\theta_j(t) - \theta_i(t)),
\end{equation}
where the oscillator $i$ has the phase angle $\theta_i$ and the natural frequency $\omega_i$, $\lambda$ represents the overall coupling strength between the oscillators and $A_{ij}$, $i, j \in \{1, \ldots, N\}$, are the elements of the adjacency matrix, i.e., $A_{ij} = 1$ if oscillators $i$ and $j$ are connected and $A_{ij} = 0$, otherwise. The degree of coherence in the system is quantified by the order parameter, 
$R e^{i\psi(t)}=  \frac{1}{N} \sum_{j=1}^N e^{i \theta_j}$ and synchronization emerges when $\lambda > \lambda_c$. For uncorrelated networks, the critical coupling has been shown to be $\lambda_c = \frac{2 \langle k \rangle}{\pi g(\bar{\omega}) \langle k^2 \rangle}  $~\cite{Ichinomiya04}. Thus, synchronization is expected to depend on the distribution of the natural frequencies, $g(\omega)$, and the network structure~\cite{Rodrigues016}. We define the state of an oscillator by considering a dummy variable as $Y_i=1$ if $\lim_{t \rightarrow \infty}|\dot{\theta}_i(t) - \Omega(t)| < 1 / \sqrt{N}$ and 0 otherwise, where $\Omega (t) = \dot{\psi}(t)$ is the mean-field frequency.   

\begin{figure}[!t] 
\begin{center}
\includegraphics[width=.9\linewidth]{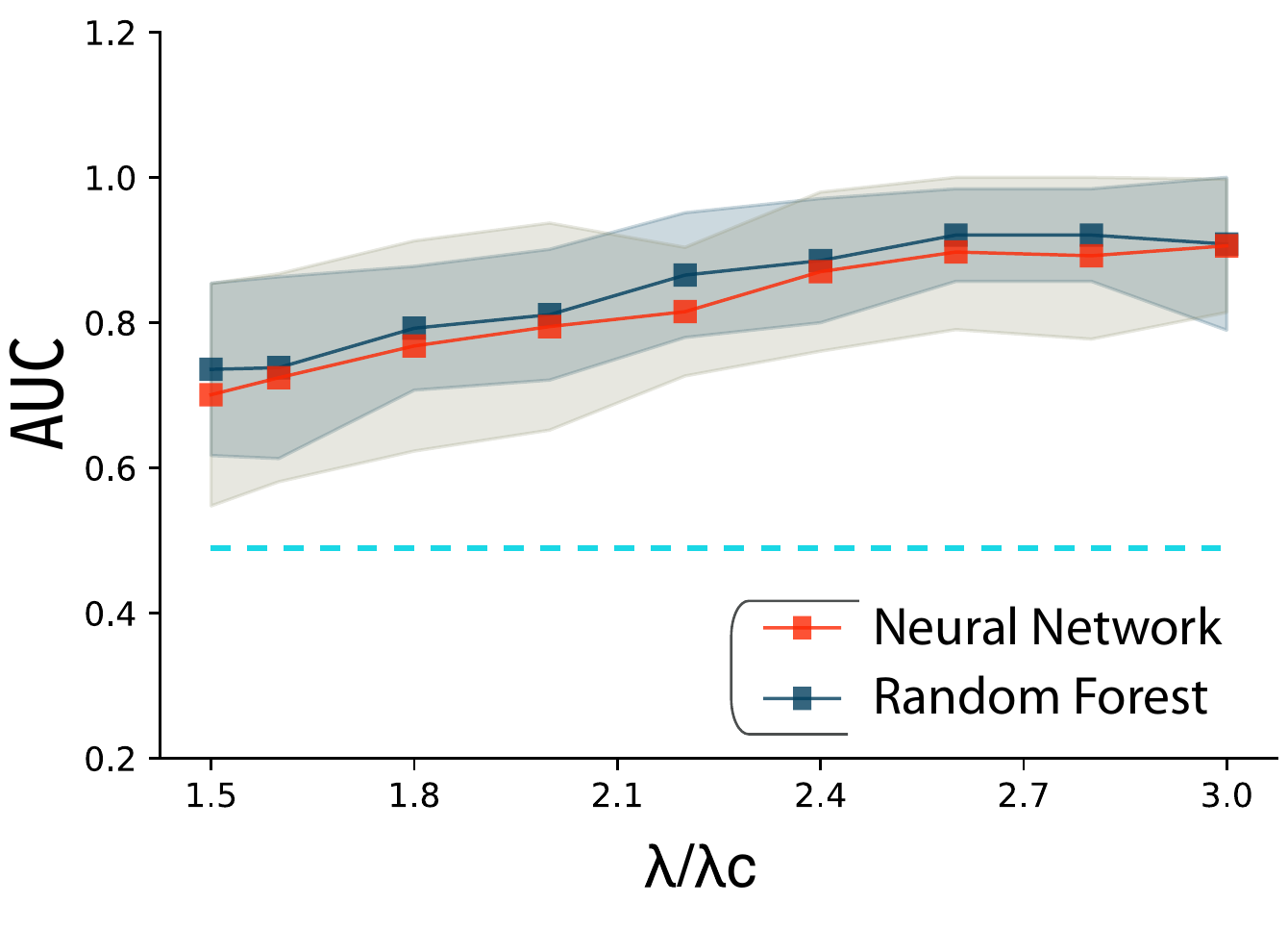}
\end{center}
\caption{AUC in function of the coupling strength for the neural network of \emph{C. elegans}. The random forest and artificial neural networks are considered to predict the state of the oscillators. The dashed line indicates the AUC for a random classification of oscillators.}
\label{fig:ROC}
\end{figure}

\begin{figure}[!t]
\begin{center}
\includegraphics[width=1\columnwidth]{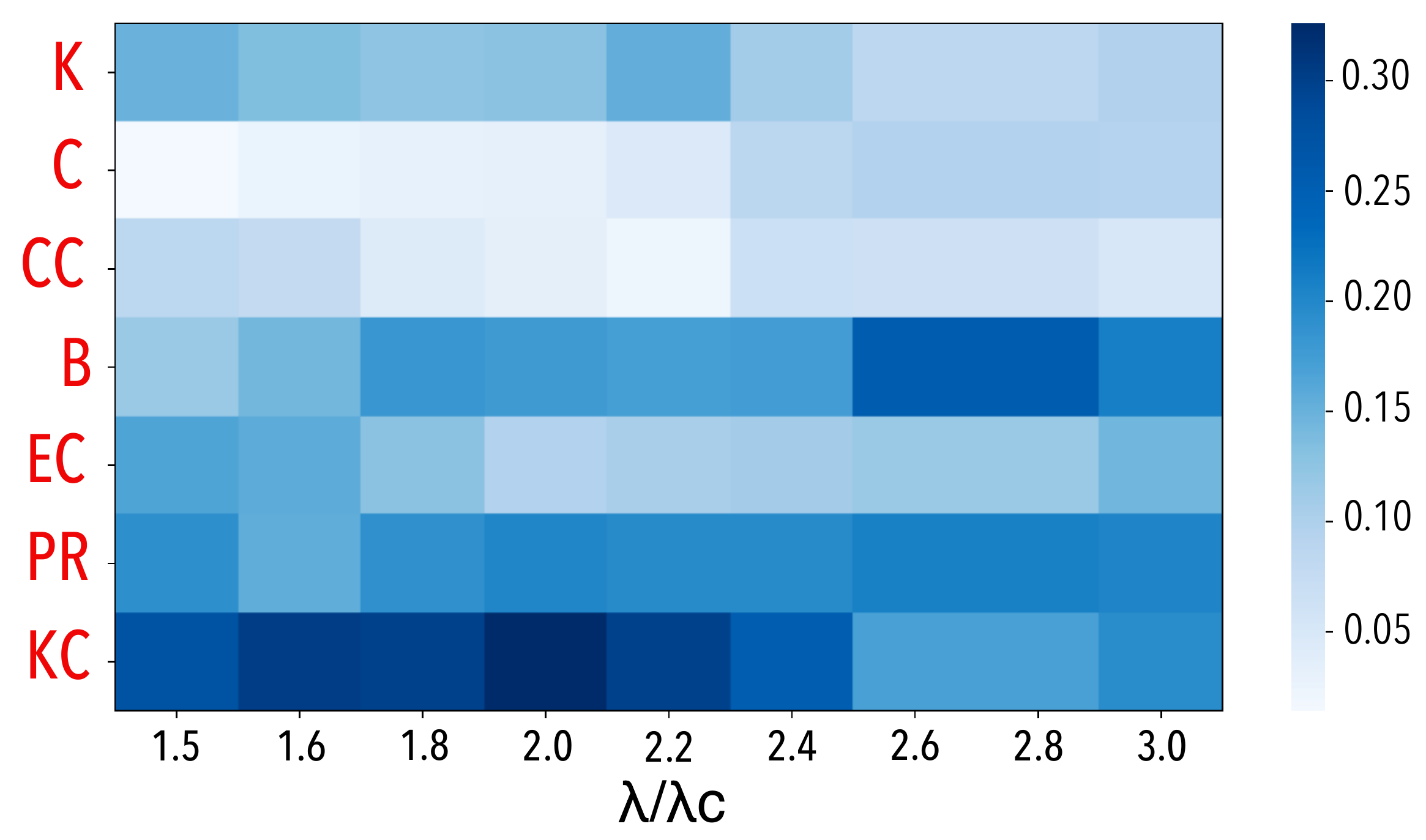}
\end{center}
\caption{Measures' importance quantified by the random forest algorithm in the neural network of the worm \emph{C. elegans}. The intensity of color represents the importance evaluated through the random forest algorithm.}
\label{fig:importance_class}
\end{figure}

We predict the variable $Y_i$ using the neural network of the worm \emph{C. elegans} ($N = 297$, $M = 2359$)~\cite{watts1998collective}. To simulate the KM, we consider the Runge-Kutta integration method with a time step $dt=0.01$ and a total simulation time $T = 400$, ignoring data for $t \in [0,100]$. The value of the coupling strength determines the fraction of synchronized oscillators. Note that in this case, for very small values of $\lambda$, most oscillators are drifting, whereas for large $\lambda$, almost all of them are synchronized. Thus, depending on the value of $\lambda$, we can have imbalanced classes and therefore the accuracy of the classifier cannot be measured by the fraction of correct classifications. We have calculated the Receiver Operating Characteristic (ROC) curve and the corresponding Area Under the Curve (AUC) measure~\cite{Mehta018}. For a classifier with no predictive power, i.e., a random guessing, $AUC = 0.5$ and the ROC curve follows the diagonal. Figure~\ref{fig:ROC} shows the AUC curve for the \emph{C. elegans} network considering different coupling strengths. As before, the random forest and the multilayer perceptron provide similar levels of accuracy, being the AUC larger than 0.8 for most of the strengths. This confirms that the synchronization level of the whole system is predictable if we know the topology and the degree of synchronization of a fraction of oscillators. Moreover, as previously done, we can quantify the importance of each network measure on the synchronization dynamics. In figure~\ref{fig:importance_class}, we present the results of this analysis. The higher the weight of a given feature in the random forest approach, the more important is the feature to predict the local synchronization. Interestingly enough, the ranking of the metrics' importance depends on the level of synchronization. For small coupling strengths, the $k$-core plays a key role. However, as we increase the coupling, measures related to the large scale organization, like betweenness centrality, play a more important role on the prediction of the nodes' states.

\begin{figure}[!t]
\begin{center}
\includegraphics[width=1\columnwidth]{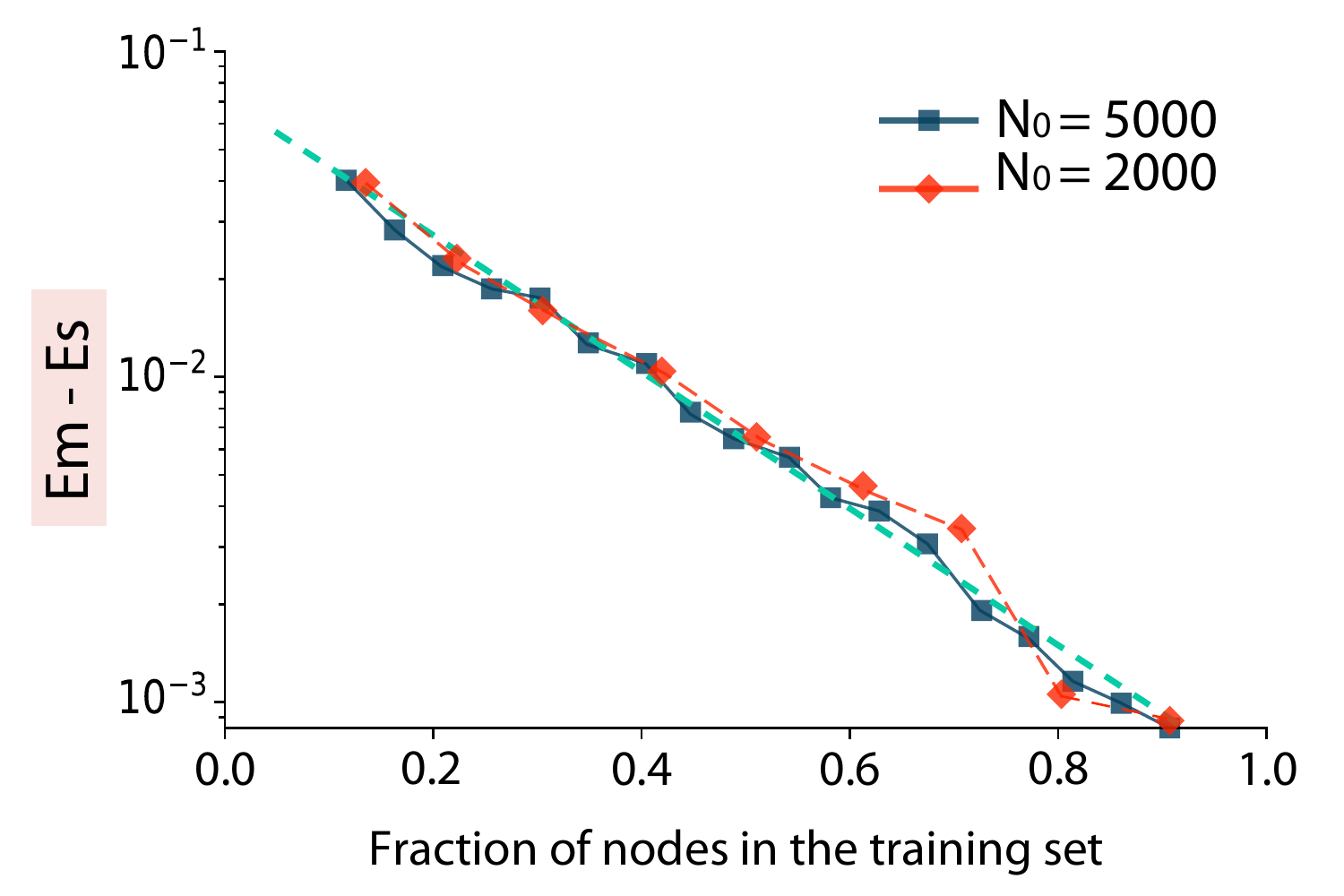}
\end{center}
\caption{Difference between the mean absolute error calculated for network generated by the Barab\'{a}si-Albert model ($E_m$) and by sampling the original network in which we make the prediction. $N_0$ is the size of the original network. See text for details.}
\label{fig:prediction-small}
\end{figure}

Before concluding, let us show that our approach can also be applied when the structure of the network is not available, but the generative model of the network is known. To illustrate this point, we consider the disease dynamics and instead of sampling a real network, we use networks of different sizes generated using the Barab\'{a}si-Albert (BA) recipe to train the ML model. More specifically, smaller networks are used as samples to train the algorithm and perform the prediction on the larger network. We measure the accuracy of the prediction through the mean-absolute error, defined as $\mathrm{MAE} = \frac{1}{N} \sum_{i=1}^N |Y_i - \hat{Y}_i|$, where $\hat{Y}_i = \hat{f}(\mathbf{X_i})$ is the predicted value of $Y_i$ by the regression model. This measure provides the same scale as the one measured in the epidemic process. Figure~\ref{fig:prediction-small} presents the results for BA networks of $N_0=1000$ and $N_0=5000$ nodes. We calculate the error by selecting the samples from the network at random ($E_s$), and by generating networks from the BA model ($E_m$). Note that the networks generated by the model has the same number of nodes as used when sampling from the original network. As we can see, the difference between these errors decreases exponentially with the network size, revealing that as the size of the training network approaches $N_0$, the network sampling is similar to the random selection of nodes. Moreover, the difference between the errors is very small, indicating that the prediction is accurate even when very small networks are used to train the ML algorithm. Thus, if we know the model that generates a network, we can use it to train the ML model and perform the prediction in very large networks.

In summary, we have shown that it is possible to estimate dynamical outcomes in complex networks by extracting features from a small number of nodes. The methodology presented is general enough as to be accurate for two different dynamics, the spreading of diseases and the synchronization of coupled oscillators. Our results are relevant for the analysis of dynamical processes on networks and pave the way for predicting dynamics from structural knowledge. For instance, the approach presented here might enable the simulation of the dynamics of very large systems whose structure is given by a generative model by simulating smaller system sizes. Additionally, recent developments in related areas such as network inference could provide further hints to refine methods for sampling the training set. These issues, among others, will be explored in future works. Hence, our methodology opens a way to assess the dynamics of complex networks using machine learning algorithms, a direction that has already been explored in other areas of Complex Systems and Physics~\cite{Zdeborov017, Carrasquilla017,Byers017,Pathak018}. 

\paragraph*{Acknowledgments.} FAR acknowledges support from CNPq (grant 307974/2013-8), FAPESP (grants 17/50144-0 and 16/25682-5), and The Leverhulme Trust for the Visiting Professorship provided. TP acknowledges FAPESP (grant 2016/23827-6). This researched is also supported by FAPESP (grant 2015/50122-0) and DFG-GRTK (grant 1740/2). YM acknowledges partial support from the Government of Aragon, Spain through grant E36-17R (FENOL), by MINECO and FEDER funds (FIS2017-87519-P) and from Intesa Sanpaolo Innovation Center.

\bibliographystyle{apsrev}
\bibliography{references}

\begin{thebibliography}{26}
\expandafter\ifx\csname natexlab\endcsname\relax\def\natexlab#1{#1}\fi
\expandafter\ifx\csname bibnamefont\endcsname\relax
  \def\bibnamefont#1{#1}\fi
\expandafter\ifx\csname bibfnamefont\endcsname\relax
  \def\bibfnamefont#1{#1}\fi
\expandafter\ifx\csname citenamefont\endcsname\relax
  \def\citenamefont#1{#1}\fi
\expandafter\ifx\csname url\endcsname\relax
  \def\url#1{\texttt{#1}}\fi
\expandafter\ifx\csname urlprefix\endcsname\relax\def\urlprefix{URL }\fi
\providecommand{\bibinfo}[2]{#2}
\providecommand{\eprint}[2][]{\url{#2}}

\bibitem[{\citenamefont{Boccaletti et~al.}(2006)\citenamefont{Boccaletti,
  Latora, Moreno, Chavez, and Hwang}}]{Boccaletti06}
\bibinfo{author}{\bibfnamefont{S.}~\bibnamefont{Boccaletti}},
  \bibinfo{author}{\bibfnamefont{V.}~\bibnamefont{Latora}},
  \bibinfo{author}{\bibfnamefont{Y.}~\bibnamefont{Moreno}},
  \bibinfo{author}{\bibfnamefont{M.}~\bibnamefont{Chavez}}, \bibnamefont{and}
  \bibinfo{author}{\bibfnamefont{D.-U.} \bibnamefont{Hwang}},
  \bibinfo{journal}{Physics Reports} \textbf{\bibinfo{volume}{424}},
  \bibinfo{pages}{175} (\bibinfo{year}{2006}).

\bibitem[{\citenamefont{Barrat et~al.}(2008)\citenamefont{Barrat, Barthelemy,
  and Vespignani}}]{Barrat08}
\bibinfo{author}{\bibfnamefont{A.}~\bibnamefont{Barrat}},
  \bibinfo{author}{\bibfnamefont{M.}~\bibnamefont{Barthelemy}},
  \bibnamefont{and}
  \bibinfo{author}{\bibfnamefont{A.}~\bibnamefont{Vespignani}},
  \emph{\bibinfo{title}{Dynamical processes on complex networks}}
  (\bibinfo{publisher}{Cambridge University Press}, \bibinfo{year}{2008}).

\bibitem[{\citenamefont{Pastor-Satorras
  et~al.}(2015{\natexlab{a}})\citenamefont{Pastor-Satorras, Castellano,
  Van~Mieghem, and Vespignani}}]{Pastor015}
\bibinfo{author}{\bibfnamefont{R.}~\bibnamefont{Pastor-Satorras}},
  \bibinfo{author}{\bibfnamefont{C.}~\bibnamefont{Castellano}},
  \bibinfo{author}{\bibfnamefont{P.}~\bibnamefont{Van~Mieghem}},
  \bibnamefont{and}
  \bibinfo{author}{\bibfnamefont{A.}~\bibnamefont{Vespignani}},
  \bibinfo{journal}{Reviews of Modern Physics} \textbf{\bibinfo{volume}{87}},
  \bibinfo{pages}{925} (\bibinfo{year}{2015}{\natexlab{a}}).

\bibitem[{\citenamefont{Rodrigues et~al.}(2016)\citenamefont{Rodrigues, Peron,
  Ji, and Kurths}}]{Rodrigues016}
\bibinfo{author}{\bibfnamefont{F.~A.} \bibnamefont{Rodrigues}},
  \bibinfo{author}{\bibfnamefont{T.~K.~D.} \bibnamefont{Peron}},
  \bibinfo{author}{\bibfnamefont{P.}~\bibnamefont{Ji}}, \bibnamefont{and}
  \bibinfo{author}{\bibfnamefont{J.}~\bibnamefont{Kurths}},
  \bibinfo{journal}{Physics Reports} \textbf{\bibinfo{volume}{610}},
  \bibinfo{pages}{1} (\bibinfo{year}{2016}).

\bibitem[{\citenamefont{de~Arruda et~al.}(2018)\citenamefont{de~Arruda,
  Rodrigues, and Moreno}}]{Arruda018}
\bibinfo{author}{\bibfnamefont{G.~F.} \bibnamefont{de~Arruda}},
  \bibinfo{author}{\bibfnamefont{F.~A.} \bibnamefont{Rodrigues}},
  \bibnamefont{and} \bibinfo{author}{\bibfnamefont{Y.}~\bibnamefont{Moreno}},
  \bibinfo{journal}{Physics Reports} \textbf{\bibinfo{volume}{756}},
  \bibinfo{pages}{1 } (\bibinfo{year}{2018}).

\bibitem[{\citenamefont{Pastor-Satorras
  et~al.}(2015{\natexlab{b}})\citenamefont{Pastor-Satorras, Castellano,
  Van~Mieghem, and Vespignani}}]{Satorras015}
\bibinfo{author}{\bibfnamefont{R.}~\bibnamefont{Pastor-Satorras}},
  \bibinfo{author}{\bibfnamefont{C.}~\bibnamefont{Castellano}},
  \bibinfo{author}{\bibfnamefont{P.}~\bibnamefont{Van~Mieghem}},
  \bibnamefont{and}
  \bibinfo{author}{\bibfnamefont{A.}~\bibnamefont{Vespignani}},
  \bibinfo{journal}{Reviews of Modern Physics} \textbf{\bibinfo{volume}{87}},
  \bibinfo{pages}{925} (\bibinfo{year}{2015}{\natexlab{b}}).

\bibitem[{\citenamefont{Kitsak et~al.}(2010)\citenamefont{Kitsak, Gallos,
  Havlin, Liljeros, Muchnik, Stanley, and Makse}}]{Kitsak010}
\bibinfo{author}{\bibfnamefont{M.}~\bibnamefont{Kitsak}},
  \bibinfo{author}{\bibfnamefont{L.~K.} \bibnamefont{Gallos}},
  \bibinfo{author}{\bibfnamefont{S.}~\bibnamefont{Havlin}},
  \bibinfo{author}{\bibfnamefont{F.}~\bibnamefont{Liljeros}},
  \bibinfo{author}{\bibfnamefont{L.}~\bibnamefont{Muchnik}},
  \bibinfo{author}{\bibfnamefont{H.~E.} \bibnamefont{Stanley}},
  \bibnamefont{and} \bibinfo{author}{\bibfnamefont{H.~A.} \bibnamefont{Makse}},
  \bibinfo{journal}{Nature Physics} \textbf{\bibinfo{volume}{6}},
  \bibinfo{pages}{888} (\bibinfo{year}{2010}).

\bibitem[{\citenamefont{de~Arruda et~al.}(2013)\citenamefont{de~Arruda, Peron,
  de~Andrade, Achcar, and Rodrigues}}]{Arruda013}
\bibinfo{author}{\bibfnamefont{G.~F.} \bibnamefont{de~Arruda}},
  \bibinfo{author}{\bibfnamefont{T.~K.~D.} \bibnamefont{Peron}},
  \bibinfo{author}{\bibfnamefont{M.~G.} \bibnamefont{de~Andrade}},
  \bibinfo{author}{\bibfnamefont{J.~A.} \bibnamefont{Achcar}},
  \bibnamefont{and} \bibinfo{author}{\bibfnamefont{F.~A.}
  \bibnamefont{Rodrigues}}, \bibinfo{journal}{Journal of Statistical Physics}
  \textbf{\bibinfo{volume}{152}}, \bibinfo{pages}{519} (\bibinfo{year}{2013}).

\bibitem[{\citenamefont{Arenas et~al.}(2008)\citenamefont{Arenas,
  D{\'\i}az-Guilera, Kurths, Moreno, and Zhou}}]{Arenas08}
\bibinfo{author}{\bibfnamefont{A.}~\bibnamefont{Arenas}},
  \bibinfo{author}{\bibfnamefont{A.}~\bibnamefont{D{\'\i}az-Guilera}},
  \bibinfo{author}{\bibfnamefont{J.}~\bibnamefont{Kurths}},
  \bibinfo{author}{\bibfnamefont{Y.}~\bibnamefont{Moreno}}, \bibnamefont{and}
  \bibinfo{author}{\bibfnamefont{C.}~\bibnamefont{Zhou}},
  \bibinfo{journal}{Physics Reports} \textbf{\bibinfo{volume}{469}},
  \bibinfo{pages}{93} (\bibinfo{year}{2008}).

\bibitem[{\citenamefont{Mehta et~al.}(2018)\citenamefont{Mehta, Bukov, Wang,
  Day, Richardson, Fisher, and Schwab}}]{Mehta018}
\bibinfo{author}{\bibfnamefont{P.}~\bibnamefont{Mehta}},
  \bibinfo{author}{\bibfnamefont{M.}~\bibnamefont{Bukov}},
  \bibinfo{author}{\bibfnamefont{C.-H.} \bibnamefont{Wang}},
  \bibinfo{author}{\bibfnamefont{A.~G.} \bibnamefont{Day}},
  \bibinfo{author}{\bibfnamefont{C.}~\bibnamefont{Richardson}},
  \bibinfo{author}{\bibfnamefont{C.~K.} \bibnamefont{Fisher}},
  \bibnamefont{and} \bibinfo{author}{\bibfnamefont{D.~J.}
  \bibnamefont{Schwab}}, \bibinfo{journal}{Preprint arXiv:1803.08823}
  (\bibinfo{year}{2018}).

\bibitem[{\citenamefont{Rodrigues}(2018)}]{Rodrigues018}
\bibinfo{author}{\bibfnamefont{F.~A.} \bibnamefont{Rodrigues}},
  \emph{\bibinfo{title}{From nonlinear dynamics to complex systems: A
  Mathematical modeling approach}} (\bibinfo{publisher}{Springer},
  \bibinfo{year}{2018}), chap. \bibinfo{chapter}{Network centrality: an
  introduction}.

\bibitem[{\citenamefont{James et~al.}(2013)\citenamefont{James, Witten, Hastie,
  and Tibshirani}}]{James013}
\bibinfo{author}{\bibfnamefont{G.}~\bibnamefont{James}},
  \bibinfo{author}{\bibfnamefont{D.}~\bibnamefont{Witten}},
  \bibinfo{author}{\bibfnamefont{T.}~\bibnamefont{Hastie}}, \bibnamefont{and}
  \bibinfo{author}{\bibfnamefont{R.}~\bibnamefont{Tibshirani}},
  \emph{\bibinfo{title}{An introduction to statistical learning}}, vol.
  \bibinfo{volume}{112} (\bibinfo{publisher}{Springer}, \bibinfo{year}{2013}).

\bibitem[{\citenamefont{Theodoridis and Koutroumbas}(2009)}]{Theodoridis}
\bibinfo{author}{\bibfnamefont{S.}~\bibnamefont{Theodoridis}} \bibnamefont{and}
  \bibinfo{author}{\bibfnamefont{K.}~\bibnamefont{Koutroumbas}},
  \emph{\bibinfo{title}{{Pattern Recognition, Fourth Edition}}}
  (\bibinfo{publisher}{{Academic Press}}, \bibinfo{year}{2009}).

\bibitem[{\citenamefont{Bishop}(2006)}]{Bishop06}
\bibinfo{author}{\bibfnamefont{C.~M.} \bibnamefont{Bishop}},
  \emph{\bibinfo{title}{Pattern recognition and machine learning}}
  (\bibinfo{publisher}{Springer}, \bibinfo{year}{2006}).

\bibitem[{\citenamefont{Theodoridis and Koutroumbas}(2008)}]{Theodoridis08}
\bibinfo{author}{\bibfnamefont{S.}~\bibnamefont{Theodoridis}} \bibnamefont{and}
  \bibinfo{author}{\bibfnamefont{K.}~\bibnamefont{Koutroumbas}},
  \emph{\bibinfo{title}{Pattern Recognition, Fourth Edition}}
  (\bibinfo{publisher}{Academic Press, Inc.}, \bibinfo{address}{Orlando, FL,
  USA}, \bibinfo{year}{2008}), \bibinfo{edition}{4th} ed.

\bibitem[{\citenamefont{Pedregosa et~al.}(2011)\citenamefont{Pedregosa,
  Varoquaux, Gramfort, Michel, Thirion, Grisel, Blondel, Prettenhofer, Weiss,
  Dubourg et~al.}}]{scikit-learn}
\bibinfo{author}{\bibfnamefont{F.}~\bibnamefont{Pedregosa}},
  \bibinfo{author}{\bibfnamefont{G.}~\bibnamefont{Varoquaux}},
  \bibinfo{author}{\bibfnamefont{A.}~\bibnamefont{Gramfort}},
  \bibinfo{author}{\bibfnamefont{V.}~\bibnamefont{Michel}},
  \bibinfo{author}{\bibfnamefont{B.}~\bibnamefont{Thirion}},
  \bibinfo{author}{\bibfnamefont{O.}~\bibnamefont{Grisel}},
  \bibinfo{author}{\bibfnamefont{M.}~\bibnamefont{Blondel}},
  \bibinfo{author}{\bibfnamefont{P.}~\bibnamefont{Prettenhofer}},
  \bibinfo{author}{\bibfnamefont{R.}~\bibnamefont{Weiss}},
  \bibinfo{author}{\bibfnamefont{V.}~\bibnamefont{Dubourg}},
  \bibnamefont{et~al.}, \bibinfo{journal}{Journal of Machine Learning Research}
  \textbf{\bibinfo{volume}{12}}, \bibinfo{pages}{2825} (\bibinfo{year}{2011}).

\bibitem[{\citenamefont{Keeling and Rohani}(2008)}]{Keeling08}
\bibinfo{author}{\bibfnamefont{M.~J.} \bibnamefont{Keeling}} \bibnamefont{and}
  \bibinfo{author}{\bibfnamefont{P.}~\bibnamefont{Rohani}},
  \emph{\bibinfo{title}{Modeling infectious diseases in humans and animals}}
  (\bibinfo{publisher}{Princeton University Press}, \bibinfo{year}{2008}).

\bibitem[{\citenamefont{De~Arruda et~al.}(2014)\citenamefont{De~Arruda,
  Barbieri, Rodr{\'\i}guez, Rodrigues, Moreno, and
  da~Fontoura~Costa}}]{Arruda014}
\bibinfo{author}{\bibfnamefont{G.~F.} \bibnamefont{De~Arruda}},
  \bibinfo{author}{\bibfnamefont{A.~L.} \bibnamefont{Barbieri}},
  \bibinfo{author}{\bibfnamefont{P.~M.} \bibnamefont{Rodr{\'\i}guez}},
  \bibinfo{author}{\bibfnamefont{F.~A.} \bibnamefont{Rodrigues}},
  \bibinfo{author}{\bibfnamefont{Y.}~\bibnamefont{Moreno}}, \bibnamefont{and}
  \bibinfo{author}{\bibfnamefont{L.}~\bibnamefont{da~Fontoura~Costa}},
  \bibinfo{journal}{Physical Review E} \textbf{\bibinfo{volume}{90}},
  \bibinfo{pages}{032812} (\bibinfo{year}{2014}).

\bibitem[{\citenamefont{G{\'o}mez et~al.}(2010)\citenamefont{G{\'o}mez, Arenas,
  Borge-Holthoefer, Meloni, and Moreno}}]{Gomez010}
\bibinfo{author}{\bibfnamefont{S.}~\bibnamefont{G{\'o}mez}},
  \bibinfo{author}{\bibfnamefont{A.}~\bibnamefont{Arenas}},
  \bibinfo{author}{\bibfnamefont{J.}~\bibnamefont{Borge-Holthoefer}},
  \bibinfo{author}{\bibfnamefont{S.}~\bibnamefont{Meloni}}, \bibnamefont{and}
  \bibinfo{author}{\bibfnamefont{Y.}~\bibnamefont{Moreno}},
  \bibinfo{journal}{EPL (Europhysics Letters)} \textbf{\bibinfo{volume}{89}},
  \bibinfo{pages}{38009} (\bibinfo{year}{2010}).

\bibitem[{\citenamefont{Rossi and Ahmed}(2015)}]{hamsterster}
\bibinfo{author}{\bibfnamefont{R.~A.} \bibnamefont{Rossi}} \bibnamefont{and}
  \bibinfo{author}{\bibfnamefont{N.~K.} \bibnamefont{Ahmed}}, in
  \emph{\bibinfo{booktitle}{AAAI}} (\bibinfo{year}{2015}),
  \urlprefix\url{http://networkrepository.com}.

\bibitem[{\citenamefont{Ichinomiya}(2004)}]{Ichinomiya04}
\bibinfo{author}{\bibfnamefont{T.}~\bibnamefont{Ichinomiya}},
  \bibinfo{journal}{Physical Review E} \textbf{\bibinfo{volume}{70}},
  \bibinfo{pages}{026116} (\bibinfo{year}{2004}).

\bibitem[{\citenamefont{Watts and Strogatz}(1998)}]{watts1998collective}
\bibinfo{author}{\bibfnamefont{D.~J.} \bibnamefont{Watts}} \bibnamefont{and}
  \bibinfo{author}{\bibfnamefont{S.~H.} \bibnamefont{Strogatz}},
  \bibinfo{journal}{nature} \textbf{\bibinfo{volume}{393}},
  \bibinfo{pages}{440} (\bibinfo{year}{1998}).

\bibitem[{\citenamefont{Zdeborov{\'a}}(2017)}]{Zdeborov017}
\bibinfo{author}{\bibfnamefont{L.}~\bibnamefont{Zdeborov{\'a}}},
  \bibinfo{journal}{Nature Physics} \textbf{\bibinfo{volume}{13}},
  \bibinfo{pages}{420} (\bibinfo{year}{2017}).

\bibitem[{\citenamefont{Carrasquilla and Melko}(2017)}]{Carrasquilla017}
\bibinfo{author}{\bibfnamefont{J.}~\bibnamefont{Carrasquilla}}
  \bibnamefont{and} \bibinfo{author}{\bibfnamefont{R.~G.} \bibnamefont{Melko}},
  \bibinfo{journal}{Nature Physics} \textbf{\bibinfo{volume}{13}},
  \bibinfo{pages}{431} (\bibinfo{year}{2017}).

\bibitem[{\citenamefont{Byers}(2017)}]{Byers017}
\bibinfo{author}{\bibfnamefont{J.}~\bibnamefont{Byers}},
  \bibinfo{journal}{Nature Physics}  (\bibinfo{year}{2017}).

\bibitem[{\citenamefont{Pathak et~al.}(2018)\citenamefont{Pathak, Hunt, Girvan,
  Lu, and Ott}}]{Pathak018}
\bibinfo{author}{\bibfnamefont{J.}~\bibnamefont{Pathak}},
  \bibinfo{author}{\bibfnamefont{B.}~\bibnamefont{Hunt}},
  \bibinfo{author}{\bibfnamefont{M.}~\bibnamefont{Girvan}},
  \bibinfo{author}{\bibfnamefont{Z.}~\bibnamefont{Lu}}, \bibnamefont{and}
  \bibinfo{author}{\bibfnamefont{E.}~\bibnamefont{Ott}},
  \bibinfo{journal}{Physical Review Letters} \textbf{\bibinfo{volume}{120}},
  \bibinfo{pages}{024102} (\bibinfo{year}{2018}).

\end{thebibliography}

\end{document}